\documentclass[prl,superscriptaddress, twocolumn]{revtex4-1}
\usepackage[T1]{fontenc}

\usepackage[english]{babel} 
\usepackage{graphicx}
\usepackage{shadow}
\usepackage{epsfig}
\usepackage{natbib}
\usepackage{dcolumn}
\usepackage{hyperref}
\usepackage{float}
\hypersetup{
backref=true,
pagebackref=true,
hidelinks
}
\begin{document}

\title{ Probing sub-eV particles with polarized Helium 3\\
\textit{Proceeding of the "Journ\'{e}es de Rencontre Jeunes Chercheurs" 2013}}
\author         {M. Guigue}
\email          {guigue@lpsc.in2p3.fr}
\affiliation{LPSC, Universit\'{e} Grenoble Alpes, CNRS/IN2P3, Grenoble INP}
\date{\today}

\begin{abstract}
Measuring the depolarization rate of a $^3$He hyperpolarized gas is a sensitive method to probe hypothetical short-range spin-dependent forces. 
A dedicated experiment is being set up at the Institute Laue Langevin in Grenoble to improve the sensitivity. 
The status of the experiment is detailed in this paper.
\end{abstract}

\maketitle

\section{Theoretical motivations}

Numerous theories beyond the Standard Model of particle physics predict the existence of new light scalar bosons \cite{Dobrescu2006} such as the Axions theory \cite{Moody1984} developed to solve the strong CP problem.
Any new scalar of mass $m_{\phi}$ could mediate a short-range monopole-dipole CP-violating interaction between two nucleons defined by the potential \cite{Moody1984}:
\begin{equation}\label{dipmonointeraction}
 V=g_s^N g_p ^N \frac{\hbar \widehat{\sigma}.\widehat{r}}{8\pi Mc}\left(\frac{m_{\phi}}{r}+\frac{1}{r^2}\right)\exp (-m_{\phi} rc/\hbar)
\end{equation}
where $\hbar\sigma/2$ is the spin of the nucleon, $M$ the mass of the polarized particle, $g_s^N$ and $ g_p^N$ the coupling constant at the vertices of polarized and unpolarized particles corresponding to a scalar and a pseudoscalar interactions. 
As a consequence, the range of such interaction is
$$\lambda=\frac{\hbar}{m_{\phi}c}$$
which makes this force macroscopically observable for light bosons.

Finding a new boson or a new short-range interaction interaction would be a important discovery in fundamental physics, since it could solve problems such as the nature of the Dark Matter (as a WISP candidate). 
This new force is thus actively searched for around the world through different kind of experiments (using torsion-balance, studying the Newton's inverse square law or looking at bouncing ultracold neutrons).

\section{ How to probe new short-range interactions with polarized Helium 3}

Consider a cell filled with polarized $^3$He immersed into a static magnetic field $B_0$. 
The natural depolarization of the gas is quantified by the longitudinal spin relaxation rate $\Gamma _1$, resulting from the main contributions:
\begin{equation}\label{eq:sum_Gamma}
\Gamma _1= \Gamma _w + \Gamma _{dd} + \Gamma _m
\end{equation}
where $\Gamma _w$ is the relaxation rate induced by collisions of polarized atoms on the cell walls, $\Gamma _{dd}$ the relaxation rate due to interparticles collisions and $\Gamma _m$ the depolarization rate due to the motion of polarized particles in an inhomogeneous magnetic field.

In our case, the axionlike interaction acts like a macroscopic pseudomagnetic field. 
The glass walls of the cell would act as a source of the pseudomagnetic field.
The induced inhomogeneities can depolarize an initially polarized gas in addition to the natural depolarization phenomena inducing $\Gamma _m$. 
The existence of such a new interaction could be revealed as an anomalous depolarization channel with a non-standard dependence on the holding magnetic field $B_0$.

\section{Polarizing Helium 3 with Tyrex}

Helium 3 is a gas which can be hypolarized (up to $80\%$ polarization) and whose polarization remains during more than 5 days under certain conditions.
In physics, $^3$He is commonly used  as a spin-filter for neutrons \cite{Andersen2006}, as a target for electron beam \cite{Rohe2006}, as a precision magnetometer \cite{Wilms1997} and as a probe for new fundamental spin-dependent interactions \cite{Petukhov2010,Fu2011,Fu2011a,Bulatowicz2013,Tullney2013}. 
In medicine and engineering, it is used for magnetic resonance imaging \cite{Safiullin2013}. 
It also has been used as a atomic zero-field magnetometer \cite{Cohen-Tannoudji1970a}.

To polarize atoms, essentially two methods are commonly used, both with optical pumping.
Since the $^3$He atoms have a nuclear polarization and so they can not be optically pumped directly in their ground state, one have to use an extra specie which can be.
In the Spin-Exchange Optical Pumping (SEOP) method, Rubidium is pumped with a $795\, \rm{nm}$ laser and then collisionally polarizes the helium atoms.

Another method is the Metastability Exchange Optical Pumping (MEOP) where an electric discharge excites the $^3$He atoms into a metastable state which can be optically pumped with a $\rm{1083\, nm}$ right circular polarized laser. 
The metastable polarized atoms then exchange their metastability with ground state helium atoms, leading to a polarized gas.
%

The $^3$He MEOP station "Tyrex" is located at the Institut Laue Langevin and is one of the best polarized $^3$He source in the world in terms of polarization and density of polarized atoms.
Since the MEOP method operates at few hundreds of millibar, a compressor is used to extract polarized atoms and to fill dedicated cells: pressures up to $4\, \rm{bars}$ at $70\%$ polarization can be reached.


\section{Status of the experiment}

\subsection{"Previous experiment"}

\begin{figure}[hbtp]
\includegraphics[scale=0.24]{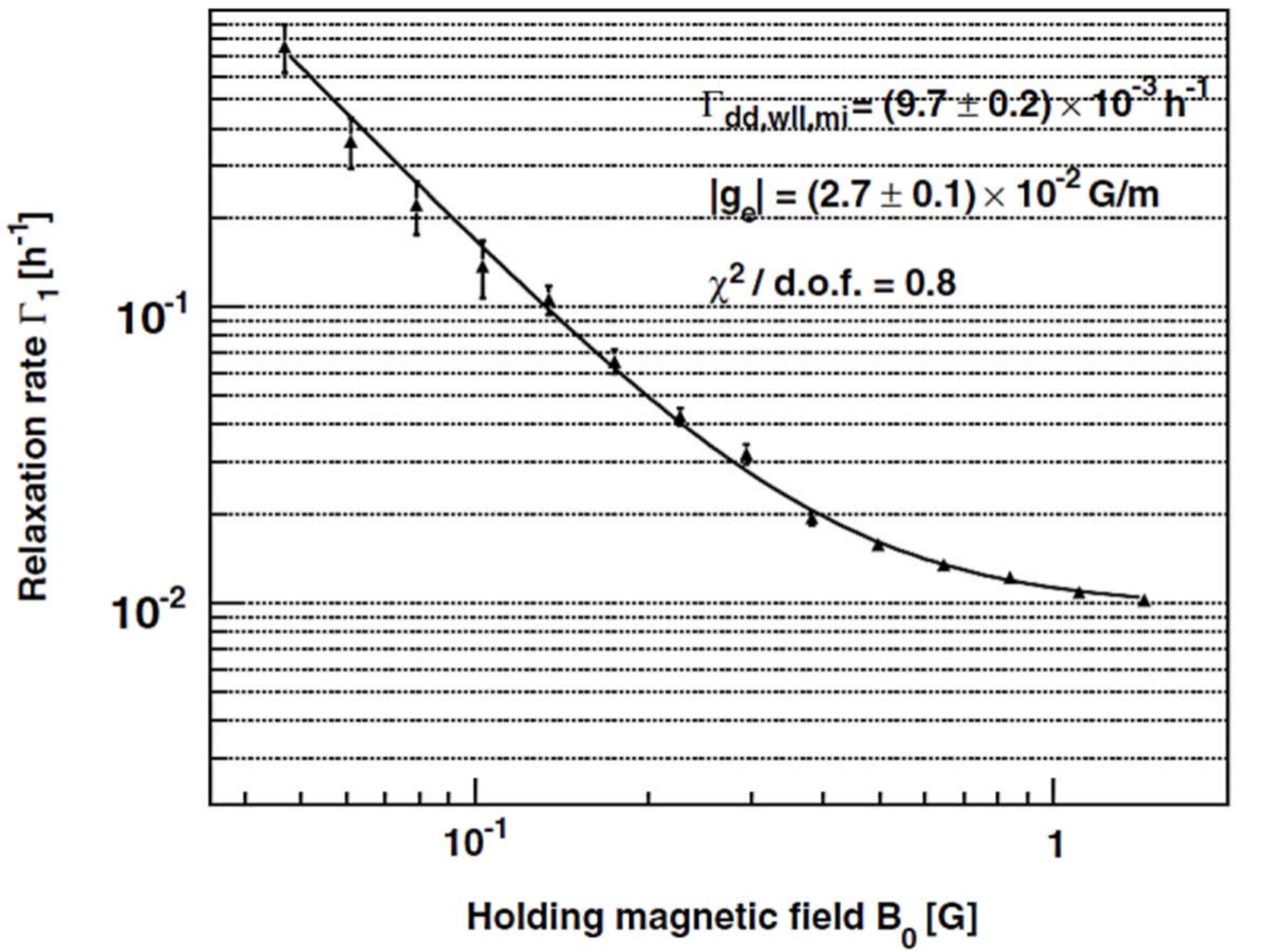}
\caption{\label{OldRelaxFigure}Relaxation rate depending on the holding magnetic field $B_0$ for the previous experiment presented in \cite{Petukhov2010} with a cylindrical cell containing $^3$He gas at $p=0.3\, \rm{bar}$ with a radius $R=5\, \rm{cm}$ and a length $L=10\, \rm{cm}$. }
\end{figure}

A first test experiment \cite{Petukhov2010} measuring the spin longitudinal depolarization rate $\Gamma_1$ as a function of the applied field $B_0$ (shown on the Figure \ref{OldRelaxFigure}) was performed in 2010 to demonstrate the sensitivity of the method. A new dedicated experiment will be set up at the Institute Laure Langevin, improving both (i) the magnetic environment of the experiment and (ii) the measurement of the decay of polarization.

In the first test experiment, (i) was  the limiting factor. 
Concerning (ii), one can improve the sensitivity using polarimetry of the polarized gas at high pressure.

\subsection{Magnetic environment}

Under certain conditions of pressure, intensity of the holding magnetic field and dynamics of particles in the cell, the longitudinal relaxation rate $\Gamma _m$ due to particles motion in the magnetic field inhomogeneities can be expressed as:
\begin{equation}
\Gamma _m = D\frac{\left\langle \left( \overrightarrow{\nabla}b_x\right) ^2 + \left( \overrightarrow{\nabla}b_y\right) ^2\right\rangle}{B_0 ^2}
\end{equation}
where $D$ is the diffusion coefficient of the $^3$He gas and $b_x$ and $b_y$ are the components of the magnetic field inhomogeneities, transversal to the holding magnetic field $B_0$.
In order to suppress the magnetic field inhomogeneity depolarization channel, one can decrease the diffusion coefficient by increasing the gas pressure.
Having a holding magnetic field as homogeneous as possible will also help to decrease $\Gamma _m$.

\begin{figure}[hbtp]
\includegraphics[scale=0.14]{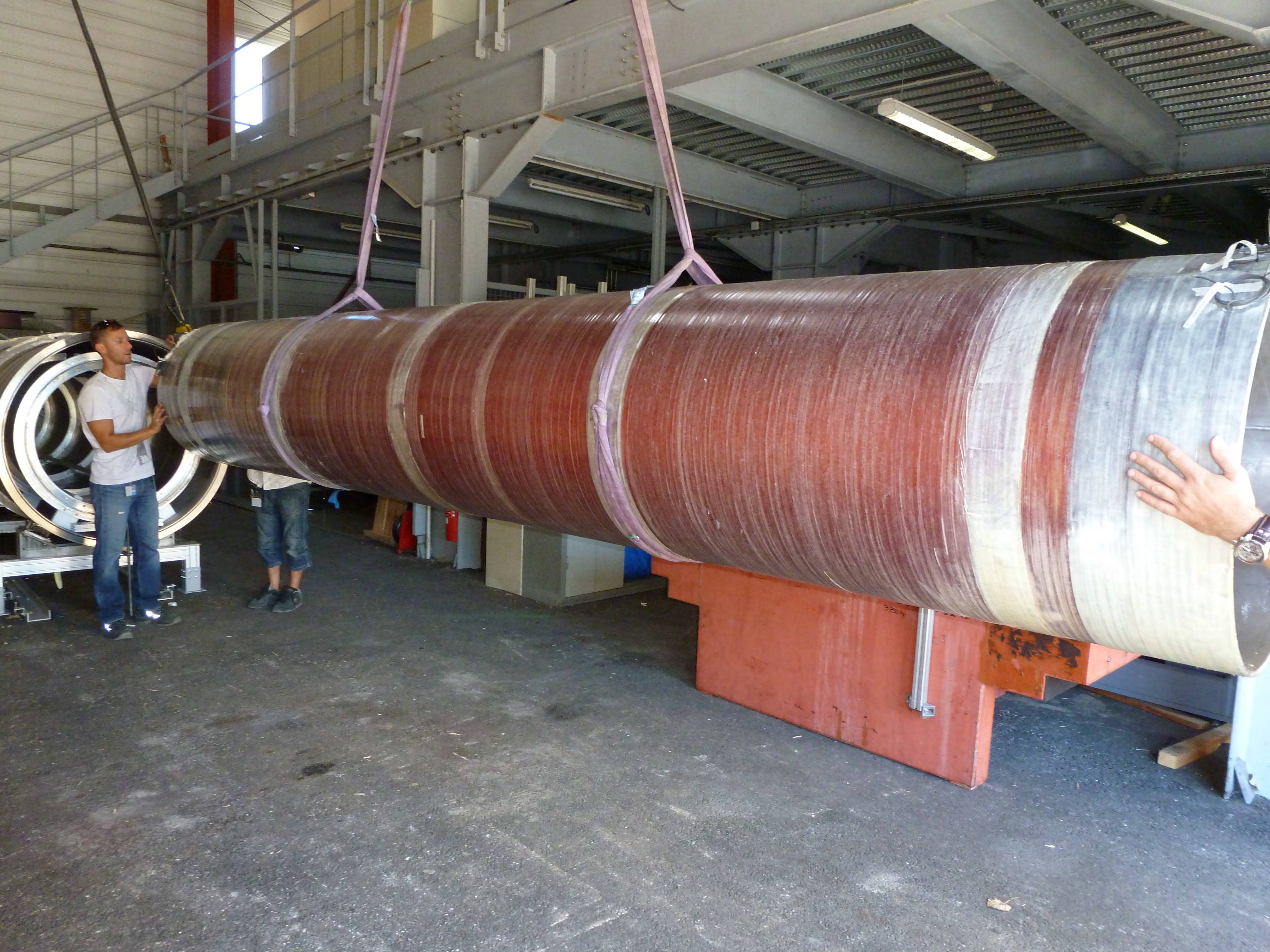}
\includegraphics[scale=0.14]{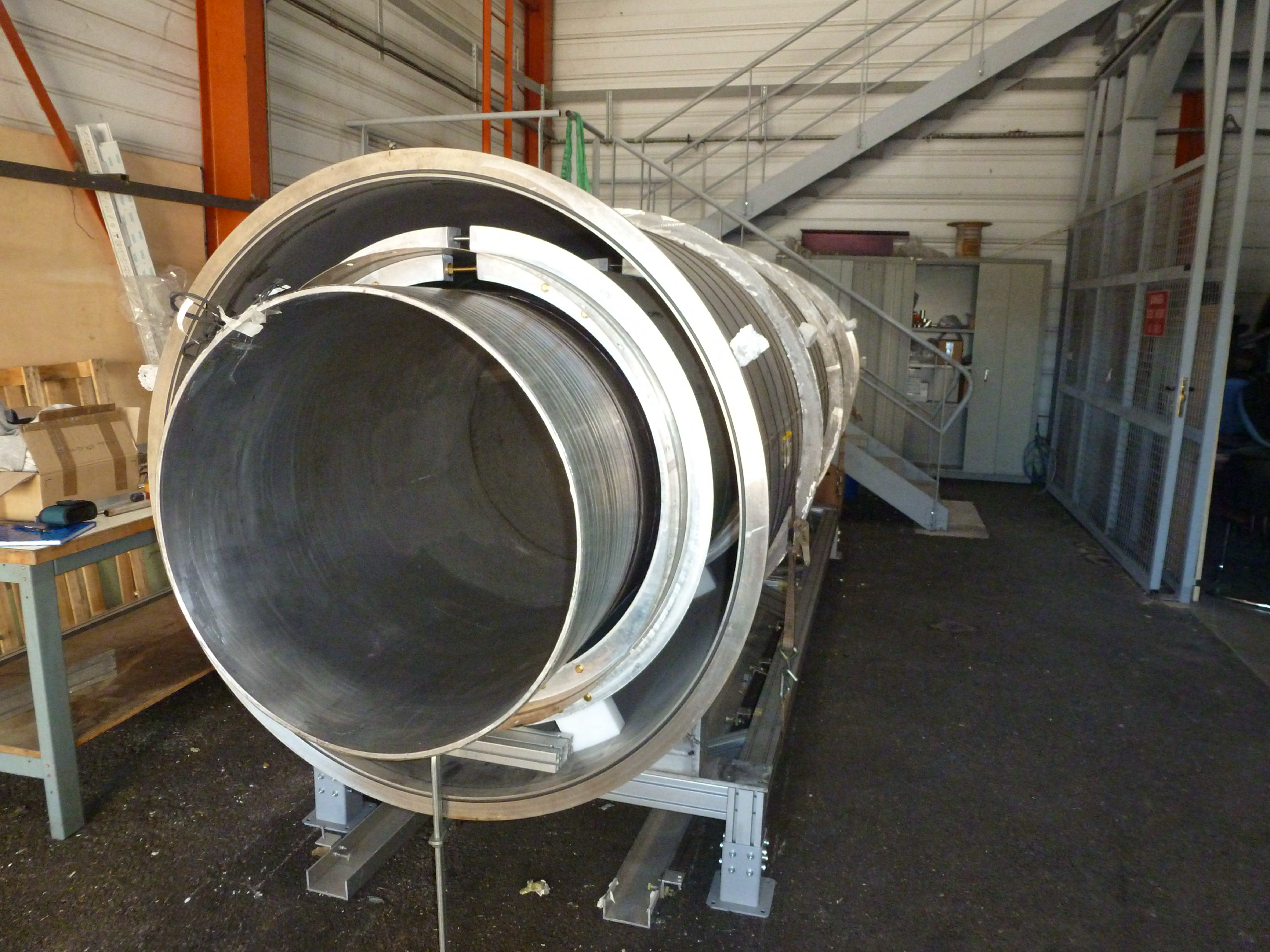}
\caption{\label{fig:experiment} Upper picture: The $5\, \rm{m}$ long copper-wired solenoid. Lower picture: the solenoid inserted into the 4 meter long $\mu$-metal magnetic shield. }
\end{figure}

The new apparatus (see Fig. \ref{fig:experiment}) is composed of a 5 meter long and 80 cm diameter solenoid which provides a very homogeneous magnetic field.
This solenoid is inserted into a $\mu$-metal magnetic shield of 96 cm diameter in order to shield the center of the solenoid from external magnetic fields.
This $4\, \rm{m}$ long tube was retrieved from "n-nbar" experiment which measured at the Institut Laue Langevin the neutron-antineutron oscillation into a $10\, \rm{nT}$ magnetic field.  

\begin{figure}[hbtp]
\center
\includegraphics[scale=0.350]{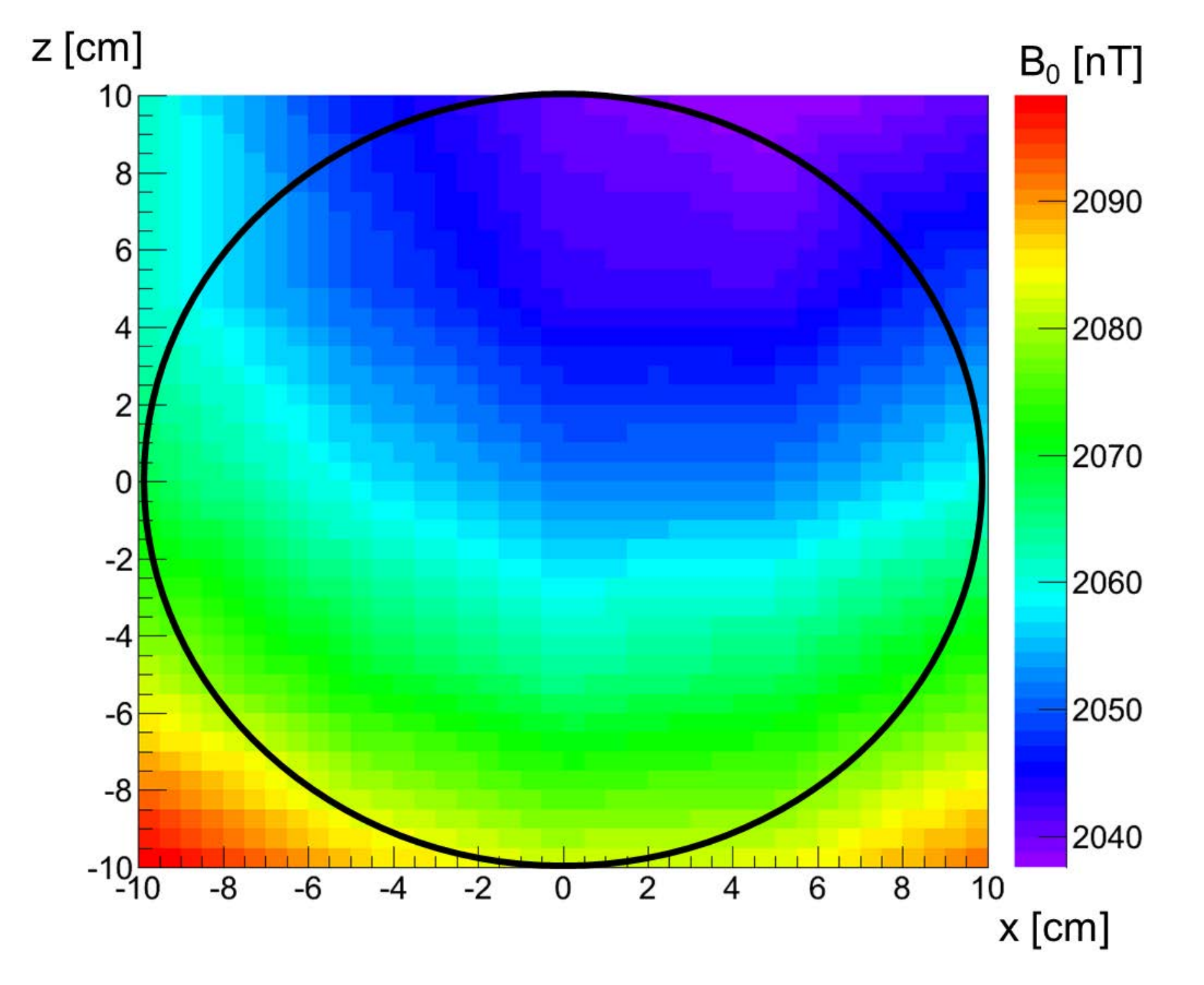}
\caption{\label{fig:carto} Longitudinal magnetic field map for a holding magnetic field of $2\, \rm{\mu T}$.
The black circle represents the place of a typical cell which will be used. }
\end{figure}

For different magnetic field $B_0$, a map of 20 $\times$ 20 $\times$ 20 $\rm{cm}^3$ has been done using a three axis fluxgate magnetometer and giving typical maps, such as the one presented in Fig. \ref{fig:carto}.
With a second-order polynomial fit, the value of the transversal gradients $g_{\perp}$ are obtained for different values of $B_0$.
Typically, $g_{\perp}$ is about $1.5\, \rm{nT/cm}$ to $2\, \rm{nT/cm}$ for magnetic field from $2\, \rm{\mu T}$ to $80\, \rm{\mu T}$.
An estimation of the expected magnetic relaxation rate $\Gamma _m$ is then calculated: the magnetic relaxation time $T_1$ is expected to be longer than $90\, \rm{h}$.
At low field ($B_0 = 3\, \rm{\mu T}$), this is a factor 100 better than the previous experiment.

This improvement will directly affect the duration of the experiment and so the precision of the relaxation rate measurement technique, which is described below.
The magnetic environment characterization is still under progress.

\subsection{Measurement of the longitudinal relaxation rate}

There are various way to measure the longitudinal relaxation rate of a polarized $^3$He gas.
The most precise but the most complex one is by measuring the transmission of an unpolarized neutrons beam through the polarized gas. 

Nuclear magnetic resonance (NMR) is a more compact and widely used technique to measure polarization.
One can measure the amplitude of the NMR response signal or can also measure the frequency shift, both proportional to the polarization.
But those techniques are also less precise.

\begin{figure}[hbtp]
\center
\includegraphics[scale=0.7]{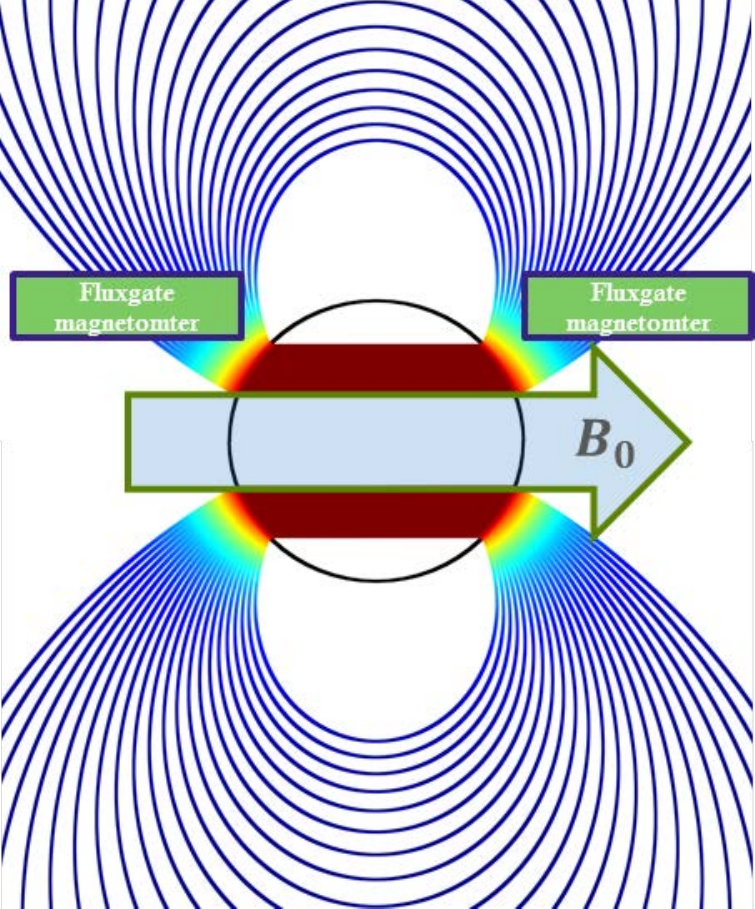}
\caption{\label{fig:polarimetry} Scheme of the polarimetry technique. A spherical cell filled with polarized helium 3 is immersed into a holding magnetic field $B_0$.
Two fluxgate magnetometers measure the magnetic field generated by the cell polarized gas.}
\end{figure}

A compromise of compactness and precision can be achieved with a direct polarimetry technique \cite{Cohen-Tannoudji1970a,Wilms1997}: it consists in measuring the magnetic field generated by the gas itself at high pressure.
This magnetic field can be of order of $100\, \rm{nT}$ which is easy to measure.
Moreover, by placing wisely two fluxgate magnetometers (see Fig. \ref{fig:polarimetry}) and applying spin-flips to invert polarization and measure offsets, the longitudinal relaxation rate can be easily and precisely measured.
The expected relative precision is of the order of $10^{-4}$ which is 100 times better than with the NMR resonance technique.

\subsection{Relaxation theory for depolarizing short range interaction}

In the case of a new spin-dependent interaction defined with Eq. (\ref{dipmonointeraction}), the nucleons of the cell walls generates a pseudo-magnetic:
\begin{eqnarray}\label{defn:magField-inhom}
b(\overrightarrow{r})&=&\frac{\hbar\lambda}{2\gamma m_n}Ng_sg_p \left( 1-e^{-d/\lambda}\right) e^{-r/{\lambda}}
\end{eqnarray}
with $r$ is the distance of the polarized particle to the closest point on the cell walls, $g_s^N$ and $g_p^N$ the coupling constants and $\lambda$ the interaction range. 
$m_n$ is the nucleon mass, $N$ the nucleon density and $d$ the thickness of the wall cell (supposed uniform).
Our zone of interest corresponds to interaction ranges $\lambda$ from $10^{-4}\, \rm{m}$ to $10^{-6}\, \rm{m}$ which are very small compared with the cell characteristic size ($L\approx 10\, \rm{cm}$).
In the previous experiment \cite{Petukhov2010}, a general formula for longitudinal relaxation rate induced by such a short range interaction was derived for a 1D problem, valid for every range of interaction.

%

For our experimental conditions ($\phi _L = \gamma B_0 L^2/D \gg 1\gg \phi _{\lambda}$), a new formula applicable for all shapes of cells was found and simplifies  for a spherical cell into:
\begin{equation}
\Gamma _{1,NF} \approx \sqrt{\frac{2}{\gamma B_0}}\frac{\lambda ^2 \gamma ^2 b_a^2}{L}
\end{equation}
which has a behavior with $B_0$ very different from $\Gamma _m$.
This different behavior will permit us to extract a constraint on the new short-range depolarizing interaction.

\section{Expectations and conclusion}

\begin{figure}[hbtp]
\center
\includegraphics[scale=0.4]{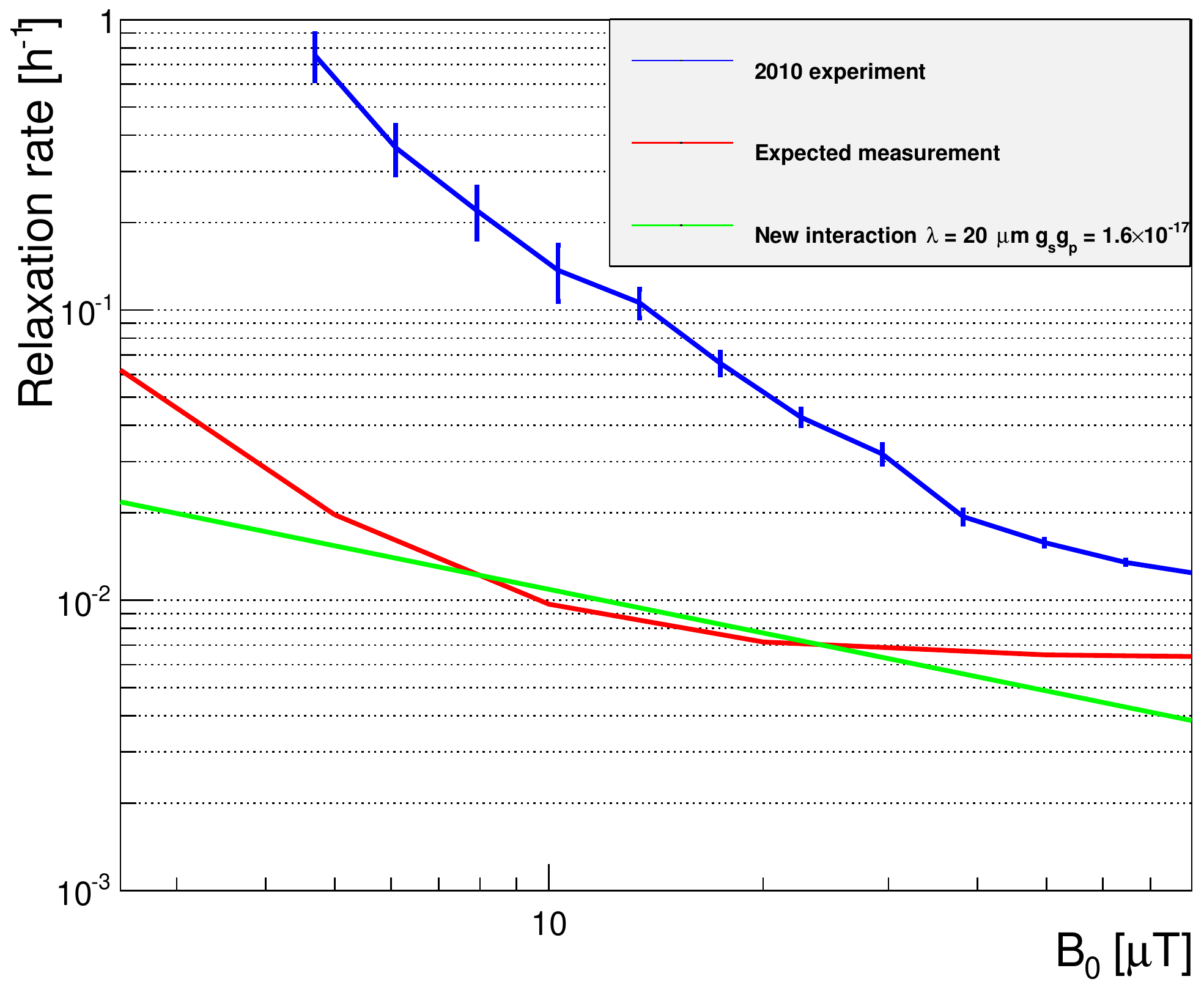}
\caption{\label{fig:GammaExp_vs_B0} Relaxation rate as a function of the magnitude of the holding magnetic field.
Blue, measurement of the previous experiment; orange, expected measurement; green, relaxation rate induced by a short range magnetic field described by Eq. (\ref{defn:magField-inhom}) for $g_s^N g_p^N = 7\times 10^{-18}$ and $\lambda = 2\times 10^{-5}\, \rm{m}$.}
\end{figure}

An estimation of the longitudinal relaxation rate $\Gamma _1$ was performed from Eq. (\ref{eq:sum_Gamma}) and is presented on Fig. \ref{fig:GammaExp_vs_B0}.
The behavior in $1/B_0^2$ of the relaxation rate due to the external magnetic field inhomogeneties is canceled by a factor 100, compared with the previous experiment.
A typical longitudinal relaxation rate due to new short-interaction depolarization channel is also plotted. It corresponds to the last point excluded by the previous experiment \cite{Petukhov2010}: it will be totally excluded by the new experiment.

\begin{figure}[hbtp]
\center
\includegraphics[scale=0.4]{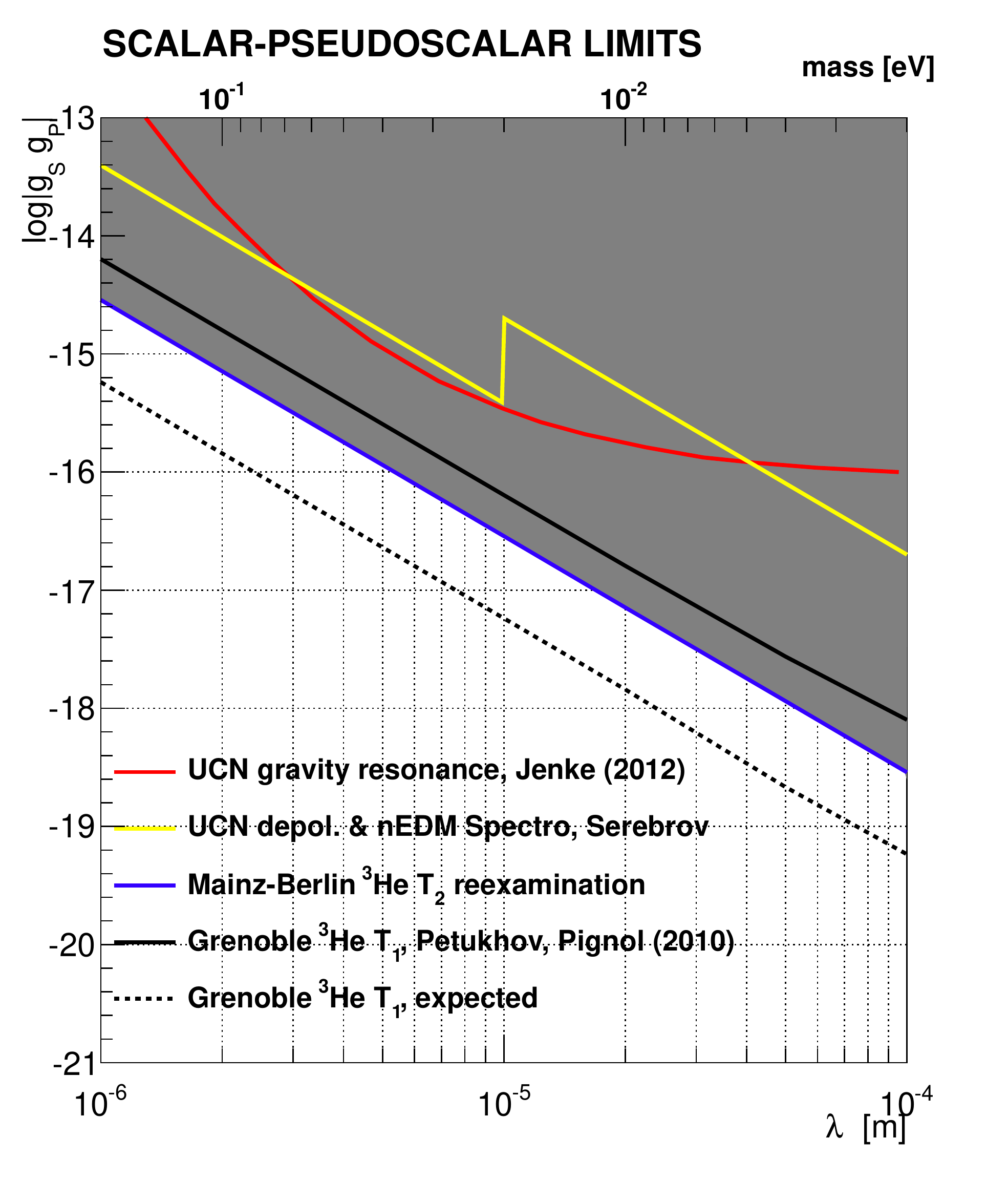}
\caption{\label{fig:exclusion_plot} Constraints on the coupling constant product of axion-like particles to nucleons $g_s^Ng_p^N$ as a function of the range $\lambda$ of the macroscopic interaction.
Yellow, combination of results from UCN precession and depolarization \cite{Serebrov2009,Serebrov2010}; red, from UCN gravitational levels \cite{Jenke2012}; bold black line, from $^3$He $T_1$ measurement \cite{Petukhov2010}; blue, $^3$He $T_2$ reexamination \cite{Petukhov2010}; dashed line, expected constraint from upgraded setup (present work). }
\end{figure}

From the expected $\Gamma _1$, one can extract the expected exclusion plot of the coupling constants product $g_s^Ng_p^N$ of this new interaction which is shown on Fig. \ref{fig:exclusion_plot}.
We expect an improvement of a factor 10 compared with the previous experiment.

\end{document}